\documentclass[12pt]{JHEP3}

 \newcommand{\be}{\begin{equation}}
 \newcommand{\ee}{\end{equation}}
 \newcommand{\bea}{\begin{eqnarray}}
 \newcommand{\eea}{\end{eqnarray}}
 \newcommand{\nn}{\nonumber}
 
 \newcommand{\rd}{\partial}

 \title{Near the horizon of $5D$ black rings}
 \author{Farhang Loran, Hesam Soltanpanahi\\
  Department of  Physics, Isfahan University of Technology,  \\
  Isfahan 84156-83111,  Iran\\
  E-mail: \email{loran@cc.iut.ac.ir, h\_soltanpanahi@ph.iut.ac.ir }}
 \abstract{For the five dimensional $\mathcal{N}=2$ black rings, we
 study the supersymmetry enhancement and identify the global supergroup
 of the near horizon geometry.
 We show that the global part of the supergroup is $OSp(4^*|2)\times U(1)$
 which is similar to the small black string.
 We show that results obtained by applying the entropy function formalism, the
 c-extremization approach and the Brown-Henneaux method to the black ring solution
 are in agreement with the microscopic entropy calculation.}

\begin{document}
 \section{Introduction and Summary}\label{int}
 Five dimensional supergravity is interesting from several points of view.
 Such a supergravity can be constructed by compactifying the eleven dimensional supergravity,
 on some six dimensional manifolds e.g. $CY_3$ or $T^6$.
 There are several supersymmetric
 solutions for the five dimensional supergravity
 that preserve either one half or all of the supersymmetry \cite{GGHPR}.
 These solutions contain three kinds of black objects which are
 half-BPS known as black holes, black strings and black rings
 which their near horizon geometries  are AdS$_2\times S^3$,
 AdS$_3\times S^2$ and AdS$_2\times S^1\times S^2$ respectively.
 Each of these solutions has a specific charge configuration.
 A black hole has only electric charges, a black string has only
 magnetic charges, while a black ring has both electric and magnetic
 charges.
 A nice review on these black solutions of ${\cal N}=2$ five dimensional supergravity is \cite{CDKL2}.

 Black ring is the first example of a
 black object with a non-spherical horizon topology  and asymptotically
 flat geometry which carries  angular momentum along the $S^1$ direction \cite{ER1}.
 Furthermore, the existence of
 this solution implies that the black hole uniqueness theorems can not
 be extended to five dimensions, except in the static case
 \cite{GIS}. The generalization of the uniqueness of black holes to five dimensions is studied in
 \cite{astefani2}, where
 it is shown that the dipole charge appears in
 the first law of thermodynamics in the same manner as a global charge.
 Therefore there exist black objects with the same {\em global} charges
 but with different horizon topologies.
 Some other  developments are listed in \cite{EEMR0}-\cite{CGMS}.
 For a good review on black ring see  \cite{ER}.

 In this paper we study some features of the supersymmetric large black
 rings in the five dimensional ${\cal N} = 2$
 supergravity which have the non-zero classical horizon area.
 Large black rings are half-BPS and in the near horizon
 limit they exhibit supersymmetry enhancement \cite{KL1}.
 We want to investigate the symmetry  of the near horizon geometry of the supersymmetric
 black rings.
 For this purpose, we note that ${\cal N} = 2n$  supergravity in five dimensions with $8n$ real supercharges
 has an $Sp(2n)$ R-symmetry group with the supersymmetry parameter
 $\varepsilon^i$, $i=1,\cdots,2n$, transforming as $\textbf{2n}$ representation.
 Using this fact, one can solve for the  supersymmetry spinor and calculate the  global part of the superalgebra.
 Doing so, we show that  the global part of the superalgebra is
 $OSp(4^*|2)\times U(1)$,
 which is similar to the small black string obtained in \cite{strominger}.

 The most important reason for investigating the supergroup
 of the near horizon geometry of the black objects is the AdS/CFT correspondence.
 AdS$_2$/CFT$_1$ correspondence is not well-defined yet in
 contrast to the higher dimensional cases (see for example \cite{S1}).
 Motivated by this phenomenon, the symmetry of the near horizon geometry of the small black hole solutions
 of ${\cal N}=2,4$ supergravity in five dimensions is studied in \cite{AAEM}.
 Lapan {\em et al} in \cite{strominger}, study the symmetry of the near horizon geometry
 of small black string solutions to
 investigate AdS$_3$/CFT$_2$ correspondence, which in principle gives
 some information about AdS$_2$/CFT$_1$ via dimensional reduction.
 Some recent results on  the  AdS$_3$/CFT$_2$ correspondence and the small black strings  can be found
 in \cite{alishahs}.
 It seems that for studying AdS$_2$/CFT$_1$  from AdS$_3$/CFT$_2$, the black ring
 is a better starting point than the black string since the
 fibration of $S^1$ over AdS$_2$ is explicit.\footnote{The same  property of black ring gives an opportunity to study the relation
 between $4D$ black holes and $5D$ black rings \cite{EEMR}.}
 Thus our study of the near horizon physics of the black ring might shed a new light on this subject.

 An important feature of supersymmetric black objects is the attractor
 mechanism.
 The attractor mechanism  determines the value of the scalar fields  near the
 horizon independent of their asymptotic values,
 and also implies the  enhancement of supersymmetry near the
 horizon \cite{FKS}.
 Attractor mechanism as reformulated by Sen, which is called the ``entropy
 function formalism",  can be used to calculate the entropy of black holes
 with AdS$_2\times X$ near horizon geometry in diverse dimensions \cite{sen}.
 In \cite{GJ,CP,CDKL} the entropy function formalism is applied to black rings.
 As mentioned above, the near horizon geometry of the black ring solution is AdS$_2\times S^1\times S^2$,
 where $S^1$ is fibred nontrivially over AdS$_2$.
 This phenomenon as well as the Chern-Simons term in the five dimensional supergravity
 frustrates a direct application of  the entropy function formalism in this
 case.\footnote{ The  difficulty in incorporating the Chern-Simons
  term into the entropy function formalism is studied in \cite{xerxes}. }
 In fact there are two problems
 in applying the entropy function method to black rings.
 First, in the Wald formula \cite{visser} there is a derivative of the Lagrangian density with respect to the Riemann tensor components
 which for  AdS$_2\times X$ near horizon geometry has only one independent component.\footnote{Interestingly, for the D1-D5-P black holes,
 where a similar problem is encountered, a generalization of the entropy function formalism is found in \cite{ghodsi},
 which can not be applied in the black ring problem.}
 But in the case of the black ring near horizon geometry,
 the  Riemann tensor has four independent components since  $S^1$ is fibred non-trivially  over
 AdS$_2$.\footnote{One encounters a Similar problem in the case of four dimensional spinning black holes.
 Applying entropy function formalism for spinning black holes is studied in \cite{astefanesi3}   }
 Second,  the Chern-Simons term in the Lagrangian density is not gauge
 invariant,\footnote{The Chern-Simons action is gauge invariant up to a boundary term.}
 while in the entropy function formalism the gauge invariance of the
 Lagrangian density is assumed.
 We study the entropy function formalism for the black ring  and
 explain how both of these problems can be resolved by dimensional reduction along the $S^1$.
 By such a dimensional reduction, the near horizon geometry reduces to
 AdS$_2\times S^2$,
 which has only one relevant independent Riemann tensor component,
 and the Chern-Simons term becomes a sum of gauge invariant terms.

 In \cite{KLc}, Kraus and Larsen  introduced the c-extremization approach
 for obtaining the spacetime central charge of black objects with AdS$_3\times Y$ near horizon geometry  in a simple way.
 Although, the c-extremization is introduced for black objects with a globally AdS$_3$
 component of the near horizon geometry,
 we show that by applying this method to the black ring
 which horizon geometry locally looks like AdS$_3\times S^2$,
 one obtains results which are in agreement with the outcome of the entropy function formalism
 and microscopic calculations of the black ring entropy  \cite{MSW,BK,CGMS,ER}.

 We recalculate the microscopic entropy by the  Kerr/CFT correspondence \cite{GHSS},
 which is intrinsically a generalization of Brown-Henneaux approach \cite{BH} to AdS/CFT correspondence  \cite{C}.
 Choosing an appropriate boundary condition we show that
 the asymptotic symmetry group of the near horizon of supersymmetric black ring
 contains a Virasoro algebra.
 The corresponding central charge equals  the
 c-extremization result.
 By defining the Frolov-Thorne temperature \cite{FT} and using the Cardy formula we
 calculate the CFT entropy and show that it equals  the
 Bekenstein-Hawking entropy.

 The main results of this work are that in five dimensional ${\cal N}=2$ supergravity
 the global part of the near horizon supergroup of the large black ring is $OSp(4^*|2)\times U(1)$.
 At the leading order, the entropy function,
 c-extremization and Brown-Henneaux approaches are in agreement with each other and
 with the microscopic results obtained in \cite{MSW,BK,CGMS,ER}.

 The paper is organized as follows.
 In  section \ref{black ring} we review black ring solution of five dimensional ${\cal N}=2$ supergravity and its near horizon geometry.
 In section \ref{enhance} we show the supersymmetry enhancement near the
 horizon of black ring and determine the global part of the
 superalgebra.
 In section \ref{physics} we apply  the entropy function,
 c-extremization and Brown-Henneaux formalisms for large black rings
 where we show that the macroscopic and microscopic entropies of black
 rings are equal to each other.
 In appendix \ref{Ap}  Killing vectors of AdS$_2\times S^1$
 component of the black ring near horizon  geometry, used in section \ref{enhance} are given.

 \section{$\mathcal{N}=2$ $5D$ black rings}\label{black ring}
 In this section we briefly review the $\mathcal{N}=2$ $5D$ black
 ring solution in superconformal formalism.
 In this approach the symmetry group of supergravity is enlarged to superconformal group
 which can be reduced to the initial model by imposing a suitable gauge fixing condition.
 The supersymmetry variations of field content are independent of the
 Lagrangian and one can consequently apply these variations at any level of
 corrections.
 \subsection{Basic setup}\label{black ring-1}

 The field content of superconformal gravity are arranged in Weyl,
 vector and hypermultiplets.
 The bosonic fields of  Weyl multiplet are the vielbein $e^a_\mu$, an
 auxiliary 2-form field $v_{ab}$ and an auxiliary scalar field $D$.
 The bosonic part of each vector multiplet contains a 1-form gauge field
 $A^{I}$ and a scalar field $X^I$, where $I=1,\cdots,n_v$ labels the
 gauge group. The hypermultiplet contains scalar fields $\mathcal{A}^i_\alpha$, where $i=1,2$ is
 the $SU(2)$ doublet index and $\alpha=1,\cdots,2n$ refers to
 $\textit{USp(2n)}$ group.

 In the off-shell formalism the bosonic part of the action of ${\cal N}=2$ supergravity in five
 dimensions at the leading order
 is \cite{HOT}
 \be
 I=\frac{1}{16\pi G_5}\int d^5x\sqrt{|g|} \mathcal{L}_0,
 \ee
 in which
 \bea
 \mathcal{L}_0&=&\rd_a\mathcal{A}^i_\alpha\rd^a\mathcal{A}^\alpha_i+(2\nu+\mathcal{A}^2)\frac{D}{4}+(2\nu-3\mathcal{A}^2)\frac{R}{8}+
 (6\nu-\mathcal{A}^2)\frac{v^2}{2}+2\nu_IF^I_{ab}v^{ab}\nn\\
 &+&\frac{1}{4}\nu_{IJ}(F^{I}_{ab}F^{J\hspace{1mm}ab}+2\rd_aX^I\rd^aX^J)+\frac{e^{-1}}{24}C_{IJK}\epsilon^{abcde}A_a^IF^J_{bc}F^K_{de}.\label{action}
 \eea
 $\mathcal{A}^2=\mathcal{A}^i_{\alpha}\mathcal{A}_i^{\alpha}$,
 $v^2=v_{ab}v^{ab}$ and
 \be
 \nu=\frac{1}{6}C_{IJK}X^IX^JX^K,\hspace{5mm}\nu_I=\frac{1}{2}C_{IJK}X^JX^K,\hspace{5mm}\nu_{IJ}=C_{IJK}X^K,
 \ee
 where $C_{IJK}$ are the intersection numbers of the internal space.
 The fermion fields are the gravitino $\psi^i_\mu$ and the auxiliary Majorana spinor $\chi^i$ which are in the Weyl multiplet,
 the gaugino $\Omega^{Ii}$  in the vector multiplet and hyperino $\zeta^\alpha$ in the hypermultiplet.

 As we are interested in supersymmetric bosonic solutions, in which fermion fields are set to zero and the solution is invariant under supersymmetry
 variations, we concentrate on the study the bosonic terms of the  supersymmetry variations of
 fermions which are given as follows\footnote{Here $\gamma_{a_1a_2\cdots
 a_m}=\frac{1}{m!}\gamma_{[a_1}\gamma_{a_2}\cdots\gamma_{a_m]}$ which
 is antisymmetric in all indices. Also the covariant curvature
 $\hat{R}^{ij}_{\mu\nu}$ is defined by
 $\hat{R}^{ij}_{\mu\nu}=2\rd_{[\mu}V_{\nu]}^{ij}-2V^i_{[\mu~k}V_{\nu]}^{kj}+$ {\em fermionic terms},
 where $V_\mu^{ij}$ is a boson in the Weyl multiplet which is in \textbf{3} of the $SU(2)$.
 For the solution we are going to consider, this term vanishes.}
 \be
 \begin{array}{l}
 \delta\psi^i_\mu=\mathcal{D}_\mu\varepsilon^i+\frac{1}{2}v^{ab}\gamma_{\mu
 ab}\varepsilon^i-\gamma_\mu\eta^i,\\\\
 \delta\chi^i=D\varepsilon^i-2\gamma^c\gamma^{ab}\hat{\mathcal{D}}_av_{bc}\varepsilon^i+\gamma^{ab}\hat{R}_{ab}(V)^i_j\varepsilon^i
 -2\gamma^a\varepsilon^i\epsilon_{abcde}v^{bc}v^{de}+4\gamma^{ab}v_{ab}\eta^i,\\\\
 \delta\Omega^{Ii}=-\frac{1}{4}\gamma^{ab}F_{ab}^I\varepsilon^i-\frac{1}{2}\gamma^a\rd_aX^I\varepsilon^i-X^I\eta^i,\\\\
 \delta\zeta^\alpha=\gamma^a\rd_a\mathcal{A}^\alpha_i-\gamma^{ab}v_{ab}\varepsilon^i\mathcal{A}^\alpha_i+3\mathcal{A}^\alpha_i\eta^i,\label{sv}
 \end{array}
 \ee
 where $\delta\equiv\bar{\epsilon}^i\textbf{Q}_i+\bar{\eta}^i\textbf{S}_i+{\xi}^a_K\textbf{K}_a$
 \footnote{$\textbf{Q}_i$ is the generator of ${\cal N}=2$ supersymmetry,
 $\textbf{S}_i$ is the generator of conformal supersymmetry
 and $\textbf{K}_a$ are special conformal boost generators of superconformal algebra \cite{HOT}.}
 and the covariant derivatives are defined by
 \bea
 &&{\cal D}_\mu\varepsilon^i=\left(\rd_\mu+\frac{1}{4}\omega_\mu^{~ab}+\frac{1}{2}b_\mu\right)-V_{\mu~j}^{~i}\varepsilon^j,\\
 &&\hat{\mathcal{D}}_\mu v_{ab}=\left(\mathcal{D}_\mu-b_\mu\right)v_{ab}=\rd_\mu v_{ab}+2\omega^{~c}_{[a}v_{b]c}-b_\mu
 v_{ab},
 \eea
 in which $b_\mu$ is a real boson in the Weyl multiplet and is $SU(2)$ singlet \cite{HOT}.

 There is a well-known gauge to fix the conformal invariance of the off-shell formalism and
 reduce the superconformal symmetry to the standard symmetries of
 five dimensional $\mathcal{N}=2$ supergravity,
 \be
 \mathcal{A}^2=-2,\hspace{5mm}b_\mu=0,\hspace{5mm}V_\mu^{ij}=0.\label{gauge}
 \ee
 In this gauge the last equation of (\ref{sv})
 gives $\eta^i$ in terms of $\varepsilon^i$ as,
 \be
 \eta^i=\frac{1}{3}\gamma^{ab}v_{ab}\varepsilon^i.\label{eta}
 \ee
 In the gauge (\ref{gauge}) and also after solving the equation of motion of the auxiliary fields $D$ and $v_{ab}$, the
 Lagrangian density (\ref{action}) reduces to the standard form of the bosonic
 part of $\mathcal{N}=2$ supergravity in five dimensions,
 \be
 \mathcal{L}_0=R-\frac{1}{2}G_{IJ}F^I_{ab}F^{J
 ab}-G_{IJ}\rd_aX^I\rd^aX^J+\frac{e^{-1}}{24}C_{IJK}A^I_aF^J_{bc}F^K_{de}\epsilon^{abcde},\label{on-shell}
 \ee
 where
 \be
 G_{IJ}=-\frac{1}{2}\rd_I\rd_J(\ln \nu)=\frac{1}{2}(\nu_I\nu_J-\nu_{IJ}),
 \ee
 and the  supersymmetry variations (\ref{sv}) simplify as
 \be
 \begin{array}{l}
 \delta\psi^i_\mu=\left(\mathcal{D}_\mu+\frac{1}{2}v^{ab}\gamma_{\mu
 ab}-\frac{1}{3}\gamma_\mu\gamma^{ab}v_{ab}\right)\varepsilon^i,\\\\
 \delta\chi^i=\left(D-2\gamma^c\gamma^{ab}\mathcal{D}_av_{bc}
 -2\gamma^a\epsilon_{abcde}v^{bc}v^{de}+\frac{4}{3}(\gamma^{ab}v_{ab})^2\right)\varepsilon^i,\\\\
 \delta\Omega^{Ii}=\left(-\frac{1}{4}\gamma^{ab}F_{ab}^I-\frac{1}{2}\gamma^a\rd_aX^I-\frac{1}{3}X^I\gamma^{ab}v_{ab}\right)\varepsilon^i,
 \end{array}
 \label{1}
 \ee
 where we have used (\ref{eta}).
 In \S \ref{enhance} we use these results for investigating
 supersymmetry enhancement near the horizon of black ring solution.
 \subsection{Black ring solutions}\label{black ring-2}
 The five dimensional $\mathcal{N}=2$ supergravity have several half-BPS
 black hole, black string and black ring solutions. Here
 we review the large black ring solutions following \cite{EEMR1}.
 These solutions have both electric $Q^I$ and magnetic $p^I$ charges.
 From eleven dimensional supergravity point of view, these
 charges correspond to the $M2$ and
 $M5$-branes respectively wrapping nontrivial cycles of the internal space.
 For simplicity we study the $U(1)^3$ solution which is the most symmetric solution.
 The M-theory configuration corresponding to this solution consists  of three M2-branes and three
 M5-branes oriented as \cite{EEMR1}
 \be
 \begin{array}{cccccccccc}
 Q_1&M2&:&1&2&-&-&-&-&-\\
 Q_2&M2&:&-&-&3&4&-&-&-\\
 Q_3&M2&:&-&-&-&-&5&6&-\\
 p_1&M5&:&-&-&3&4&5&6&\psi\\
 p_2&M5&:&1&2&-&-&5&6&\psi\\
 p_3&M5&:&1&2&3&4&-&-&\psi
 \end{array}
 \ee
 where directions $z_i$, $i=1\cdots 6$, span the internal 6-torus and $\psi$ is the ring direction of black ring.

 The $11D$ supergravity solution takes the form
 \be
 \begin{array}{l}
 ds^2_{11}=ds^2_5+X^1(dz_1^2+dz_2^2)+X^2(dz_3^2+dz_4^2)+X^3(dz_5^2+dz_6^2),\\\\
 \mathcal{A}=A^1\wedge dz_1\wedge dz_2+A^2\wedge dz_3\wedge dz_4+A^3\wedge
 dz_5\wedge dz_6,
 \end{array}
 \ee
 where $\mathcal{A}$ is the three-form potential with four-form field
 strength $\mathcal{F}=d\mathcal{A}$.

 The five dimensional solution is specified by a metric $ds_5^2$, three scalars $X^I$,
 and three one-forms $A^I$, with field strengths $F^I= dA^I$.\footnote{
 All the fields are independent from internal space and exterior derivative $d$ on $A^I$ is defined in five dimensional
 space.}
 In ring coordinates the solution is written as
 follows\footnote{for metrics, we use $(+----)$ signature.}
 \be
 \begin{array}{l}
 ds_5^2=\left(H_1H_2H_3\right)^{-2/3}(dt+\omega)^2-\left(H_1H_2H_3\right)^{1/3}d\textbf{x}_4^2,\\\\
 d\textbf{x}_4^2=\frac{R^2}{(x-y)^2}\left[(y^2-1)d\psi^2+\frac{dy^2}{y^2-1}+\frac{dx^2}{1-x^2}+(1-x^2)d\phi^2\right],\\\\
 A^I=H_I^{-1}(dt+\omega)+\frac{p^I}{2}[(1+y)d\psi+(1+x)d\phi],\\\\
 X^I=H_I^{-1}\left(H_1H_2H_3\right)^{1/3}.
 \end{array}
 \label{brs}
 \ee
 In these coordinates, $y=-\infty$ corresponds to the location of the ring, and $Q^I$
 and $p^I$ are the electric and magnetic charges respectively. The
 harmonic functions $H_I$ are defined by
 \footnote{$R=0$ reduces the black ring solution to the BMPV solution \cite{BMPV} although the limit of $R\rightarrow0$ in (\ref{H}) is singular .}
 \be
 H_1=1+\frac{Q_1-p_2p_3}{2R^2}(x-y)-\frac{p_2p_3}{4R^2}(x^2-y^2),\label{H}
 \ee
 and the same for $H_2$ and $H_3$ with cyclic permutation.
 For simplicity we choose
 \be
 Q_1=Q_2=Q_3=Q,\hspace{10mm}p_1=p_2=p_3=p.\label{qQ}
 \ee

 The one-form $\omega$ which is related to the angular momentum of the solution is $\omega=\omega_\psi d\psi+\omega_\varphi d\varphi$ with
 \be
 \begin{array}{l}
 \omega_\psi=\frac{p}{8R^2}(y^2-1)[3Q-p^2(3+x+y)]-\frac{3p}{2}(1+y),\\\\
 \omega_\varphi=\frac{p}{8R^2}(1-x^2)[3Q-p^2(3+x+y)].
 \end{array}
 \ee
 The ADM charges of this solution are given by
 \be
 \begin{array}{c}
 M=\frac{3\pi}{4G_5}Q,\\\\
 J_\psi=\frac{\pi}{8G_5}p[6R^2+3Q-p^2],\hspace{7mm}
 J_\varphi=\frac{\pi}{8G_5}p(3Q-p^2).
 \end{array}
 \label{J}
 \ee
 The coordinate ranges are
 \be
 -\infty\leq y\leq 1,\hspace{10mm}-1\leq x\leq
 1,\hspace{10mm}0\leq \psi\leq 2\pi,\hspace{10mm}0\leq \varphi\leq
 2\pi.
 \ee

 To make the above solution free of
 closed causal curves for $y\geq-\infty$, one requires that,
 \be
 2p^2L^2\equiv2\sum_{i<j}{\cal Q}_ip_i{\cal Q}_jp_j-\sum_{i}{\cal
 Q}_i^2p^2_i-2R^2p^3\sum_{i}p_i\geq0,\label{L0}
 \ee
 where
 \be
 p=(p_1p_2p_3)^{1/3},\hspace{7mm}{\cal Q}_1=Q_1-p_2p_3,\hspace{7mm}{\cal Q}_2=Q_2-p_1p_3,\hspace{7mm}{\cal
 Q}_3=Q_3-p_1p_2.
 \ee
 For (\ref{qQ}),
 \be
 L=\sqrt{3\left[\frac{(Q-p^2)^2}{4p^2}-R^2\right]},\label{LL}
 \ee
 There is a nice review on this solution
 by Emparan and Reall \cite{ER}.
 \subsection{Near horizon geometry}\label{black ring-3}
 In \cite{EEMR0,GJ} it is shown that in a comfortable coordinate system the near
 horizon limit of the black ring solution (\ref{brs}) becomes,
 \be
 ds^2=-p^2\bigg(\frac{dr^2}{4r^2}+\frac{L^2}{p^2}d\psi^2+\frac{Lr}{p}dtd\psi\bigg)-\frac{p^2}{4}(d\theta^2+\sin^2\theta
 d\phi^2),\label{met0}
 \ee
 which is the product of a locally AdS$_3$ with radius $p$ and an $S^2$
 with radius $\frac{p}{2}$.
 This metric can be written as follows,
 \be
 ds^2=\frac{p^2}{4}(r^2dt^2-\frac{dr^2}{r^2})-L^2(d\psi+\frac{p r}{2 L}dt)^2
 -\frac{p^2}{4}(d\theta^2+\sin^2\theta d\phi^2),\label{met}
 \ee
 where the range of coordinates are
 \be
 0\leq
 r\leq\infty,\hspace{10mm}0\leq\theta\leq\pi,\hspace{10mm}0\leq\phi\leq2\pi,\hspace{10mm}0\leq\psi\leq2\pi.\label{range}
 \ee
 The near horizon geometry (\ref{met}) is AdS$_2\times S^1\times S^2$ in which the  $S^1\times AdS_2$ component,
 locally looks like the BTZ black hole with radius $r_+=L$.
 Furthermore, in these coordinates  the near horizon limit of the field strengths of the gauge fields $A^I$ (\ref{brs}) are
 \be
 F_{\theta\phi}^I=-\frac{p^I}{2}\sin\theta,\label{F}
 \ee
 and the attractor values of scalars $X^I$ are
 \be
 X^I=\frac{p^I}{(\frac{1}{6}C_{IJK}p^Ip^Jp^K)^{1/3}}.\label{X}
 \ee

 As is well-known, one can in principle obtain considerable information about a black
 object by studying the corresponding near horizon geometry.
 In the next two sections we study the enhancement of supersymmetry near the horizon
 and apply the entropy function, the c-extremization and
 Brown-Henneaux formalisms to obtain the global supergroup of the near horizon
 geometry and the entropy of black ring.

 \section{Enhancement of supersymmetry}\label{enhance}
 The bosonic  supersymmetric solution of any supersymmetric
 theory should be invariant under supersymmetry variations of fermions.
 Thus both fermions and their supersymmetry variations should be equal to zero.

 For studying enhancement of supersymmetry near the horizon of any
 half-BPS black object one should check if it is possible to
 make supersymmetry variations of fermions of the theory equal to zero without imposing any additional
 condition on the Killing spinor $\varepsilon^i$.
 In our case  the supersymmetry variations of auxiliary Majorana spinor $\chi^{i}$  and gaugino
 $\Omega^{I i}$ in (\ref{1}) give no constraint on spinor
 $\varepsilon^i$, but imposes the following constraints on the
 bosonic auxiliary fields,
 \be
 F^I_{\mu\nu}=-\frac{4}{3}X^Iv_{\mu\nu},\hspace{5mm}D=\frac{8}{3}v^2\hspace{5mm} \epsilon^{abcd}{\cal D}_av_{bc}=0,\hspace{5mm}{\cal
 D}^bv_{ab}-\frac{1}{3}\epsilon_{abcd}v^{bc}v^{de}=0.\label{v}
 \ee
 Thus we only investigate the gravitino variation in $\psi^i$ (\ref{1}) for supersymmetry enhancement.

 \subsection{Killing spinor}\label{enhance-1}
 The calculations will be easier in non-coordinate basis.
 The components of vielbein  for the near horizon geometry of the black ring
 solution (\ref{met}) are
 \be
 e^{\hat{t}}=\frac{p
 r}{2}dt,\hspace{3mm}e^{\hat{r}}=\frac{p}{2r}dr,\hspace{3mm}e^{\hat{\theta}}=\frac{p}{2}d\theta,\hspace{3mm}
 e^{\hat{\phi}}=\frac{p\sin\theta}{2}d\phi,\hspace{3mm}
 e^{\hat{\psi}}=L(d\psi+\frac{pr}{2L}dt),
 \ee
 and the inverse components are given by
 \be e^{\hat{t}t}=\frac{2}{p
 r},\hspace{3mm}e^{\hat{r}r}=-\frac{2r}{p},\hspace{3mm}e^{\hat{\theta}\theta}=-\frac{2}{p},\hspace{3mm}
 e^{\hat{\phi}\phi}=-\frac{2}{p\sin\theta},\hspace{3mm}
 e^{\hat{\psi}\psi}=-\frac{1}{L},\hspace{3mm}e^{\hat{t}{\psi}}=-\frac{1}{L}.
 \ee
 By using the explicit  relation for spin connection,
 \be
 (\omega_\mu)^{a b}=\frac{1}{2}e^{\nu a}(\rd_\mu e^b_\nu -\rd_\nu
 e^b_\mu)-\frac{1}{2}e^{\nu b}(\rd_\mu e^a_\nu -\rd_\nu
 e^a_\mu)-\frac{1}{2}e^{\rho a}e^{\sigma b}(\rd_\rho e_{\sigma
 c}-\rd_\sigma e_{\rho c})e^c_\mu,
 \ee
 one can show that the non zero components of spin connections are
 \be
 (\omega_t)^{\hat{t}\hat{r}}=-\frac{r}{2},\hspace{3mm}
 (\omega_t)^{\hat{r}\hat{\psi}}=\frac{r}{2},\hspace{3mm}
 (\omega_r)^{\hat{t}\hat{\psi}}=\frac{1}{2r},\hspace{3mm}
 (\omega_\phi)^{\hat{\theta}\hat{\phi}}=\cos\theta,\hspace{3mm}
 (\omega_\psi)^{\hat{t}\hat{r}}=\frac{L}{p}.
 \ee

 Supersymmetry variation of gravitino in (\ref{1}) for our background
 (\ref{met}) and (\ref{F}) simplifies to
 \be
 \delta \psi_\mu^i=(\rd_\mu+\frac{1}{4}\omega_\mu^{~~a b}\gamma_{a b}+v_{\theta\phi}\gamma_{\mu}^{~~\theta\phi}
 -\frac{2}{3}\gamma_\mu
 \gamma^{\theta\phi}v_{\theta\phi})\varepsilon^i.\label{gv1}
 \ee
 By using the attractor value of the scalars (\ref{X})
 together with the value of field strengths (\ref{F}) and the first equality in (\ref{v})
 one obtains,
 \be
 v_{\hat{\theta}\hat{\phi}}=\frac{3}{2p}.
 \ee

 Now, setting all the components of gravitino variation (\ref{gv1}) equal to
 zero gives the following equations,
 \be
 \label{v1}
 \begin{array}{l}
 \nabla_t\varepsilon^i=\left(\rd_t+\frac{r}{4}\gamma^{\hat{t}\hat{r}}(1-\gamma^{\hat{r}\hat{\theta}\hat{\phi}})-
 \frac{r}{4}\gamma^{\hat{\psi}\hat{r}}(1-\gamma^{\hat{r}\hat{\theta}\hat{\phi}})\right)\varepsilon^i=0,\\
  \nabla_r\varepsilon^i=\left(\rd_r-\frac{1}{4r}(\gamma^{\hat{r}\hat{\theta}\hat{\phi}}+\gamma^{\hat{t}\hat{\psi}})\right)\varepsilon^i=0,\\
  \nabla_\theta\varepsilon^i=\left(\rd_\theta-\frac{1}{2}\gamma^{\hat{\phi}}\right)\varepsilon^i=0,\\
  \nabla_\phi\varepsilon^i=\left(\rd_\phi+\frac{1}{2}\cos\theta\gamma^{\hat{\theta}\hat{\phi}}+
  \frac{1}{2}\sin\theta\gamma^{\hat{\theta}}\right)\varepsilon^i=0,\\
  \nabla_\psi\varepsilon^i=\left(\rd_\psi-\frac{L}{2p}(\gamma^{\hat{t}\hat{r}}+\gamma^{\hat{\theta}\hat{\phi}\hat{\psi}})\right)\varepsilon^i=0.
\end{array}
 \ee
 One can easily show that integrability condition,
 \be
 [\nabla_\mu,\nabla_\nu]\varepsilon^i=0,
 \ee
 is automatically satisfied without imposing  any
 projection on the Killing spinor $\varepsilon^i$. Thus all the supersymmetry
 is restored near the horizon.

  Assuming
  $\gamma^{\hat{t}\hat{\psi}\hat{r}\hat{\theta}\hat{\phi}}=1$ equations (\ref{v1}) simplify
  as
  \be
  \begin{array}{l}
 \nabla_t\varepsilon^i=\left(\rd_t+\frac{r}{2}(\gamma^{\hat{t}\hat{r}}-\gamma^{\hat{\psi}\hat{r}})\right)\varepsilon^i=0,\\\\
  \nabla_r\varepsilon^i=\left(\rd_r-\frac{1}{2r}\gamma^{\hat{t}\hat{\psi}}\right)\varepsilon^i=0,\\\\
  \nabla_\theta\varepsilon^i=\left(\rd_\theta-\frac{1}{2}\gamma^{\hat{\phi}}\right)\varepsilon^i=0,\\\\
  \nabla_\phi\varepsilon^i=\left(\rd_\phi+\frac{1}{2}\cos\theta\gamma^{\hat{\theta}\hat{\phi}}+
  \frac{1}{2}\sin\theta\gamma^{\hat{\theta}}\right)\varepsilon^i=0,\\\\
  \nabla_\psi\varepsilon^i=\rd_\psi\varepsilon^i=0.
  \end{array}
 \ee
 There are two solutions corresponding to the projections
  $\gamma^{\hat{r}\hat{\theta}\hat{\phi}}\varepsilon^i_{(\pm)}=\pm
  \varepsilon^i_{(\pm)}$.
  For $\gamma^{\hat{r}\hat{\theta}\hat{\phi}}\varepsilon^i=\varepsilon^i$ the above equations simplify
  to,
  \be
  \begin{array}{l}
 \nabla_t\varepsilon^i=\rd_t\varepsilon^i=0,\\\\
  \nabla_r\varepsilon^i=(\rd_r-\frac{1}{2r})\varepsilon^i=0,\\\\
  \nabla_\theta\varepsilon^i=\left(\rd_\theta-\frac{1}{2}\gamma^{\hat{\phi}}\right)\varepsilon^i=0,\\\\
  \nabla_\phi\varepsilon^i=\left(\rd_\phi+\frac{1}{2}\cos\theta\gamma^{\hat{\theta}\hat{\phi}}+
  \frac{1}{2}\sin\theta\gamma^{\hat{\theta}}\right)\varepsilon^i=0,\\\\
  \nabla_\psi\varepsilon^i=\rd_\psi\varepsilon^i=0.
  \end{array}
  \label{v11}
 \ee
 It is easy to show that there are two solutions for these  equations,
 \be
 \varepsilon^i=\sqrt{\frac{r}{l}}e^{\frac{1}{2}\gamma^{\hat{\phi}}\theta}e^{-\frac{1}{2}\gamma^{\hat{\theta}\hat{\phi}}\phi}\varepsilon^i_0,
 \hspace{1cm}\lambda^i=l(-t+\frac{1}{r}\gamma^{\hat{t}\hat{r}})\varepsilon^i,\label{solution}
 \ee
 where $l=p/2$ is the radius of both AdS$_2$ and $S^2$ part of the near horizon
 geometry (\ref{met})
 and $\varepsilon^i_0$ is a constant spinor satisfying
 $\gamma^{\hat{r}\hat{\theta}\hat{\phi}}\varepsilon^i_0=\varepsilon^i_0$.
 The relation between two different chiralities is
 $\varepsilon^i_{0(-)}=\gamma^{\hat{t}\hat{r}}\varepsilon^i_{0(+)}$.

 It is enlightening to note that the
 Killing spinors solution (\ref{solution}) depends only on the radius of AdS$_2$ and $S^2$
 which is proportional to the cube root of the central charge,
 $c=6p^3$ (\ref{c}).

 \subsection{Near horizon superalgebra}\label{enhance-2}

 As it is shown in appendix \ref{Ap} the isometries of AdS$_2\times S^1$ are generated
 by,
 \be
 \begin{array}{ll}
 K_1=\frac{1}{2}(t^2+r^{-2})\rd_t-rt\rd_r-\frac{p}{2Lr}\rd_\psi,\hspace{5mm}&K_2=t\rd_t-r\rd_r,\\\\
 K_3=\rd_\psi,&K_4=\rd_t,\\\\
 K_5=e^{\frac{2L}{p}\psi}\left(\frac{1}{r}\rd_t+r\rd_r-\frac{p}{2L}\rd_\psi\right),&
 K_6=e^{-\frac{2L}{p}\psi}\left(\frac{1}{r}\rd_t-r\rd_r-\frac{p}{2L}\rd_\psi\right).
 \end{array}\label{K5-K6}
 \ee
  The algebra associated to these isometries is as follows
 \be
 \begin{array}{lllll}
 [K_1,K_2]=-K_1,&\ & [K_1,K_4]=-K_2,&\ & [K_2,K_4]=-K_4,\\

 [K_3,K_5]=\frac{2L}{p}K_5,&\ & [K_3,K_6]=-\frac{2L}{p}K_6,&\ & [K_5,K_6]=-\frac{p}{L}K_3.
 \end{array}
 \label{algebra}
 \ee
 Using the following redefinitions,
 \be
 \begin{array}{lll}
 pK_1\rightarrow L_{1}^+,\hspace{5mm}&\frac{2}{p}K_4\rightarrow
 L_{-1}^+,\hspace{5mm}&-K_2\rightarrow L_0^+,\\\\
 K_5\rightarrow L_{1}^-,&K_6\rightarrow L_{-1}^-,&-\frac{p}{2L}K_3\rightarrow L_0^-,
 \end{array}\label{L0-}
 \ee
 the algebra (\ref{algebra}) simplifies to,
 \be
 [L^\pm_m,L^\pm_n]=(m-n)L^\pm_{m+n},\hspace{5mm}m,n=\pm 1,0.
 \ee
 This may be identified to the left and right moving global parts of
 the CFT$_2$ which is dual to this locally AdS$_3$ geometry.
 Since the generators $K_5$ and $K_6$ defined in Eq.(\ref{K5-K6})
 are not single valued functions of $\psi\sim\psi+2\pi$ the
 $SL(2)_L$ is broken to $U(1)$ generated by $K_3$.
 Consequently the global symmetry of the near horizon geometry is $U(1)_L\times
 SL(2)_R$.

 The action of these generators on the spinors
 $\varepsilon^i$ and $\lambda^i$ (\ref{solution}) can be defined by the Lie derivative
 $\mathcal{L}_K\varepsilon^i=(K^{\mu}\mathcal{D}_\mu+\frac{1}{4}\rd_\mu
 K_\nu\gamma^{\mu\nu})\varepsilon^i$, which gives,
 \be
 \begin{array}{llll}
 L_{-1}^+\lambda^i=-\varepsilon^i,\hspace{5mm}&L_0^+\lambda^i=-\frac{1}{2}\lambda^i,
 \hspace{5mm}&L_0^+\varepsilon^i=\frac{1}{2}\varepsilon^i,\hspace{5mm}&L_{1}^+\varepsilon^i=\lambda^i,\\\\
  L_{-1}^+\varepsilon^i=0,\hspace{5mm}&L_{1}^+\varepsilon^i=0,\hspace{5mm}&L_{m}^-\varepsilon^i=0,\hspace{5mm}&L_{m}^-\lambda^i=0,\label{L}
 \end{array}
 \ee

 Considering a correspondence between $\varepsilon^i$ and
 $\lambda^i$ and the $G_{-\frac{1}{2}}$ and $G_{\frac{1}{2}}$ modes
 of supercurrent $G$ then (\ref{L}) simplify to
 \be
 [L_m^+,G_r]=(\frac{m}{2}-r)G_{m+r},\hspace{10mm}[L_m^-,G_r]=0.
 \ee

 To complete the algebra, we should also study the behavior of the Killing spinors (\ref{solution}) under the generators of $S^2$.
 These generators are
 \be
 J^3=-i\rd_\phi,\hspace{20mm}
 J^\pm=e^{\pm i\phi}(-i\rd_\theta\pm\cot\theta\rd_\phi).
 \ee
 Since $\gamma^{\hat{r}\hat{\theta}\hat{\phi}}$ and
 $\gamma^{\hat{\theta}\hat{\phi}}$ commute with each other one can
 choose
 \be
 \gamma^{\hat{\theta}\hat{\phi}}\varepsilon^i_0=\pm
 i\varepsilon^i_0,
 \ee
 which gives
 \be
 J^3\varepsilon^i=\pm\frac{1}{2}\varepsilon^i,\hspace{20mm}J^3\lambda^i=\pm\frac{1}{2}\lambda^i.
 \ee
 Thus both $\varepsilon^i$ and $\lambda^i$ are in the $\textbf{2}$
 representation of the $SU(2)$ group which is generated by $J^i$s.
 If one starts with a constant spinor which satisfies
 $\gamma^{\hat{\theta}\hat{\phi}}\varepsilon^i_0=-i\varepsilon_0$\footnote{Normalization is
 $\varepsilon^\dag_0\varepsilon_0=1/2$.}
 and define
 \be
 \xi_+=\sqrt{\frac{r}{l}}e^{\frac{1}{2}\gamma^{\hat{\phi}}\theta}e^{\frac{i}{2}\phi}\varepsilon_0,\hspace{20mm}
 \xi_-=\sqrt{\frac{r}{l}}e^{\frac{1}{2}\gamma^{\hat{\phi}}\theta}e^{-\frac{i}{2}\phi}\gamma^{\hat{\theta}}\varepsilon_0,
 \ee
 one verifies that $\xi^a$ are in the  $\textbf{2}$ representation of the $SU(2)$
 group, $J^{\pm}_0\xi^a_\pm=0$ and $J^\pm_0\xi^a_\mp=\xi^a_\pm$.

 Our results so far can be organized into symplectic-Majorana killing spinors
 \footnote{symplectic-Majorana condition is $\overline{\zeta}^i=\zeta_i^\dag \gamma^{\hat{t}}={\zeta^{i}}^TC$.}
 \bea
 \varepsilon^{1}&=&\left(\begin{array}{c} \xi_+ \\  i\xi_- \end{array}\right),
 \hspace{5mm}
 \varepsilon^{2}=\left(\begin{array}{c} -i\xi_+ \\  -\xi_- \end{array}\right),
 \hspace{5mm}
 \varepsilon^{3}=\left(\begin{array}{c} \xi_- \\   -i\xi_+ \end{array}\right),
 \hspace{5mm}
 \varepsilon^{4}=\left(\begin{array}{c} i\xi_- \\
 -\xi_+\end{array}\right),\label{sc}
 \eea
 where each $\varepsilon^{I}$ transforms as $\textbf{2}$ of $Sp(2)$ and
 corresponds to $G^{I}_{-\frac{1}{2}}$. In the same manner one can
 define
 \be
 \lambda^{I}=l(-t+\frac{1}{r}\gamma^{\hat{t}\hat{r}})\varepsilon^{I}\label{sc1}
 \ee
 which corresponds to $G^{I}_{\frac{1}{2}}$ and also transforms as $\textbf{2}$ of $Sp(2)$.

 To complete the near horizon superalgebra we need to compute the anticommutators
 of supercharges. To do this we use the supersymmetry transformations
 of the five dimensional supergravity given by \cite{HOT,FO,BCDD},
 \bea
 \{G_r^I,G^J_s\}&=&
 l\Omega_{ij}\left[(\bar{\varepsilon}^I_r)^i\gamma^\mu(\varepsilon^J_s)^j+
 (\bar{\varepsilon}^J_s)^i\gamma^\mu(\varepsilon^I_r)^j\right]\rd_\mu\nn\\
 &&+\left[(\bar{\varepsilon}^I_r)_i\gamma^{\hat{\theta}\hat{\phi}}(\varepsilon^J_s)^j+
 (\bar{\varepsilon}^J_s)_i\gamma^{\hat{\theta}\hat{\phi}}(\varepsilon^I_r)^j\right],\label{anti}
 \eea
 where $\Omega_{ij}$ is a symplectic matrix which can be used for raising
 and lowering the indices as follows,
 \be
 \chi^i=\Omega^{ij}\chi_j,\hspace{1cm}\chi_i=\chi^j\Omega_{ji},
 \ee
 and we have chosen a basis $\Omega_{1 2}=1$.
 By plugging the supercharges (\ref{sc}) and (\ref{sc1}) into
 (\ref{anti}) we derive the anticommutators of the supercharges,
 \be
 \{G^{I}_{\pm\frac{1}{2}},G^{J}_{\pm\frac{1}{2}}\}=-2\delta^{IJ}L_{\pm}^+,\label{gg1}
 \ee
 and
 \be
 \{G^{I}_{-\frac{1}{2}},G^{J}_{\frac{1}{2}}\}=
 \left(\begin{array}{c}\hspace{5mm}
 -2L_0^+\hspace{11mm}2iJ^3+i\sigma_3\hspace{9mm}2iJ^2+i\sigma_1\hspace{7mm}-2iJ^1+i\sigma_2 \\
 -2iJ^3-i\sigma_3\hspace{8mm}-2L_0^+\hspace{10mm}-2iJ^1-i\sigma_2\hspace{5mm}-2iJ^2+i\sigma_3\\
 -2iJ^2-i\sigma_1\hspace{5mm}2iJ^1+i\sigma_2\hspace{11mm}-2L_0^+\hspace{16mm}2iJ^3-i\sigma_3\hspace{2mm}\\
 \hspace{4mm}2iJ^1-i\sigma_2\hspace{6mm}2iJ^2-i\sigma_1\hspace{5mm}-2iJ^3+i\sigma_3\hspace{12mm}-2L_0^+\hspace{7mm}
 \end{array}\right),\label{gg2}
 \ee
 where $m,n=0,\pm1$, $r,s=\pm\frac{1}{2}$, $I,J=1,2,3,4$ and $\sigma_a$ are the Pauli matrices.
 We can summarize all the results to the following superalgebra,
 \bea
 &&\{G^{I}_r,G^{J}_s\}=-2\delta^{IJ}L_{r+s}^++(r-s)(M_a)^{IJ}J^{a}+(r-s)(N_A)^{IJ}T^{A},\nn\\
 &&[L_m^{+},L_n^{+}]=(m-n)L_{m+n}^{+},\hspace{8mm}[L_m^+,G_r^{I}]=\left(\frac{m}{2}-r\right)G^{I}_{m+r},\nn\\
 &&[J^{\alpha},G^{I}_r]=(t^\alpha)^{IJ}G_r^J,\hspace{18mm}[T^{A},G^{I}_r]=(N^A)^{IJ}G_r^J,\label{sual}
 \eea
 where $M_a$ and $N_A$ are the representation matrices for
 $SU(2)$ and $Sp(2)$, respectively and $T^A$ are generators of $Sp(2)$.
 Therefore the bosonic part of the global supergroup is
 \be
 U(1)_L\times SL(2)_R\times SU(2)\times Sp(2),\label{sp2}
 \ee
 while the isometry of the near horizon geometry of the black  ring
 solution is $U(1)_L\times SL(2)_R\times SU(2)$.\footnote{ The symmetries of the near horizon geometry
 of the extremal black ring and four dimensional spinning black holes are studied for example in \cite{kunduri} and \cite{astefanesi1} respectively.}
 The generators in (\ref{sual}) are null under the $U(1)_L$
 generated by $L_0^-$ given in (\ref{L0-}).
 The extra $Sp(2)$ in (\ref{sp2})
 can be identified with R-symmetry of $\mathcal{N}=2$ supergravity in
 five dimensions \cite{strominger,AAEM}.

 Searching in the literature (for example see \cite{vp}) we found that there is a
 supergroup with this bosonic part and supporting eight supercharges
 which is $D(2, 1; 1)=Osp(4^*|2)$.
 So we propose that (\ref{sual}) correspond to the $Osp(4^*|2)\times
 U(1)$.

 It is interesting to note that this supergroup is the same as the small black string
 near horizon supergroup \cite{strominger}.
 Of course, in \cite{strominger}  the superalgebra of small black string in $\mathcal{N}=4$ five dimensional
 supergravity is calculated  by embedding the  solution of $\mathcal{N}=2$ supergravity.
 For this solution, the
 supergroup of near horizon is $Osp(4^*|4)\times U(1)$,
 where the $Osp(4^*|4)$ part of superalgebra is
 \bea
 &&\{G^{I}_r,G^{J}_s\}=-2\delta^{IJ}L_{r+s}+(r-s)(t_\alpha)^{IJ}J^{\alpha}+(r-s)(\rho_A)^{IJ}R^{A},\nn\\
 &&[L_m,L_n]=(m-n)L_{m+n},\hspace{8mm}[L_m,G_r^{I}]=\left(\frac{m}{2}-r\right)G^{I}_{m+r},\nn\\
 &&[J^{\alpha},G^{I}_r]=(t^\alpha)^{IJ}G_r^J,\hspace{18mm}[R^{A},G^{I}_r]=(\rho^A)^{IJ}G_r^J,\label{N4}
 \eea
 in which $t^\alpha$ and $\rho^A$ are the representation matrices for $SU(2)$ and $Sp(4)$ respectively and
 $R^A$ are the generators of $Sp(4)$.
 In \cite{AAEM}, it was shown that this global part of superalgebra in ${\cal N}=4$ supergravity
 is reduced to $Osp(4^*|2)$ in ${\cal N}=2$.\footnote{It is interesting to note  that $Osp(4^*|2)$ factor is
 also present in the small black string \cite{strominger},
 the small black hole \cite{AAEM} and the large black ring solutions (\ref{sual}) of $\mathcal{N}=2$
 supergravity in five dimensions.}
 This result shows that in AdS/CFT analysis black ring solution
 behaves like a small black string.

 It is straightforward to repeat the above calculations when higher order corrections are
 considered, as higher order corrections only modify  $p$ and  $L$ in the metric (\ref{met}) \cite{CDKL1}-\cite{small}.
 Therefore, after adding e.g. the supersymmetric correction \cite{HOT}, the supersymmetry  is still enhanced near the horizon
 and the superalgebra does not change.

 \section{Near horizon physics }\label{physics}
 A special feature of the supersymmetric black ring is the geometry of
 the near horizon of this solution such that an AdS$_2\times
 S^1$  is locally AdS$_3$ (\ref{met0})-(\ref{met}). This special near horizon topology
 allows one to apply both the entropy function \cite{sen} and
 the c-extremization \cite{KLc} formalisms on black ring.
 We also use Brown-Henneaux approach \cite{BH} to
 calculate the CFT entropy of extremal black ring.
 \subsection{Entropy function}\label{physics-1}
 In this section we briefly review  the entropy function formalism applied to the black
 ring solution to calculate the corresponding macroscopic entropy \cite{GJ}, \cite{CP}, \cite{CDKL}.

 In the entropy function formalism  the entropy  can be found from the extremum of the entropy
 function,
 \be
 S=2\pi(e^{I}q_I-f),\label{sen}
 \ee
 in which,
 \be
 f=\int dx_H\sqrt{|g_H|}{\cal L},
 \ee
 and  $q_I$ is defined by $q_I=\frac{\rd f}{\rd e^{I}}$.
 To apply the entropy function for the near
 horizon geometry of the black ring one uses the ansatz \cite{GJ}\footnote{In this section
  following the usual conventions in both the entropy function and the c-extremization formalisms we use $(-++++)$ signature and  choose $G_5=\pi/4$.}
 \be
 \begin{array}{l}
 ds^2=v_1(-r^2dt^2+\frac{dr^2}{r^2})+v_2(d\theta^2+\sin^2\theta
 d\phi^2)+w(d\psi^2+e_0rdt)^2,\\\\
 F^I_{5rt}=e^I+a^Ie_0,\hspace{10mm}F^I_{5\theta\phi}=\frac{p^I}{2}\sin\theta,\hspace{10mm}X^I=M^I,\hspace{5mm}I=1,2,3,
 \end{array}
 \label{0-solution}
 \ee
 where $e_0$ is conjugate to the angular momentum of the ring.
 Extremizing the entropy function  (\ref{sen}) with respect to the $v_1, v_2, w, M^I$ and $N^I$
 gives,
 \be
 v_1=v_2=\frac{p^2}{4},\hspace{5mm}w=\frac{p}{2e_0},\hspace{5mm}e^{I}+e_0N^{I}=0,
 \hspace{5mm}M^{I}=\frac{p^{I}}{(\frac{1}{6}C_{IJK}p^Ip^Jp^K)^{1/3}}.\label{ex}
 \ee
 Using (\ref{ex})  and (\ref{sen}) one obtains,
 \be
 S_{\rm mac}=2\pi p^2L.\label{entropy}
 \ee
  The same result is obtained in  the off-shell formalism in \cite{CP}.

 \subsection{c-extremization}\label{physics-2}
 In \cite{KLc} Kraus and Larsen showed that for $D$-dimensional black objects with  AdS$_3\times S^{D-3}$ near horizon geometry
 one can define the
 c-function as\footnote{It was discussed in \cite{KLc} that only the bulk part of the action contributes in this definition. }
 \be
 c(l_A,l_S)=\frac{3\Omega_2\Omega_{D-3}}{32\pi
 G_{D}}l_A^3l_S^{D-3}\mathcal{L},\label{cf}
 \ee
 which extremization  with respect to the radii of AdS$_3$ and $S^2$,
 gives the average of the left and right central charges of CFT
 dual of AdS$_3$.
 As we have discussed in \S\ref{black ring} and as can also be verified using the Killing vectors derived in the appendix A,
 the near horizon geometry of the black ring solution,
  has locally an AdS$_3$ component (\ref{met0}).
 Thus one can expect that c-extremization formalism \cite{KLc} can also be
 applied for black ring solution.

 We consider the following ansatz,
 \be
 \begin{array}{l}
 ds^2=l_A^2d\Omega_{AdS_3}^2+l_S^2d\Omega_{S^2}^2,\hspace{10mm} X^I=mp^I,\\\\
 F^I_{rt}=e^I+e_0a^I,\hspace{5mm} F^I_{\theta
 \phi}=\frac{p^I}{2}\sin\theta,\hspace{5mm}v_{rt}=v_1\hspace{5mm}v_{\theta
 \phi}=v_2\sin\theta.\label{ansatz}
 \end{array}
 \ee
 After solving the equations of motion of $D, v_{ab}, m$ and $a^I$
 one finds\footnote{These results can also be derived from the supersymmetry variations of fermions (\ref{1}) \cite{CP}.}
 \be
 D=\frac{12}{p^2},\hspace{5mm}m=p^{-1},\hspace{5mm}v_2=-\frac{3}{8}p,\hspace{5mm}v_1=0,\hspace{5mm}e^{I}+e_0a^{I}=0,\label{eom}
 \ee
 where $p\equiv(\frac{1}{6}C_{IJK}p^Ip^Jp^K)^{1/3}$.\footnote{In  $U(1)^3$ supergravity
 which  is the subject of our study in this paper,  all $p^I$ are
 equal to each other and denote them by $p$.}
 By extremizing the c-function one obtains,
 \be
 l_A=2l_S=p,\hspace{10mm} c=6p^3.\label{c}
 \ee
  In the semiclassical regime  $c\gg1$, $c\sim c_L\sim c_R$,
 in which $c_{L(R)}$ is the left (right) central charge of the CFT.
 In this limit, $(c_L-c_R)$ is negligible as it is given by higher order corrections, and
  $c={1\over2}(c_L+c_R)$ is given by the c-extremization method (\ref{c}).
  Since the black ring solution in ${\cal N}=2$ five dimensional supergravity  corresponds to the $(0,4)$-CFT
  the microscopic entropy is given by the logarithm of the number of left moving
 excitations \cite{MSW,BK,CGMS,ER},\footnote{We are using  the conventions of \cite{MSW} for left and right moving central charges.}
 \be
 S_{\rm mic}=2\pi\sqrt{\frac{c\hat{q}_0}{6}},\label{cardy}
 \ee
 where \cite{ER}
 \bea
 \hat{q}_0&=&\frac{p_1p_2p_3}{4}+\frac{1}{2}(\frac{Q_1Q_2}{p_3}+\frac{Q_2Q_3}{p_1}+\frac{Q_3Q_1}{p_2})\\
 &+&\frac{1}{4p_1p_2p_3}[(p_1Q_1)^2+(p_2Q_2)^2+(p_3Q_3)^2]-J_\psi\nn\\
 &=&pL^2.\label{l-q0}
 \eea
  Thus, at the leading order, the result of the c-extremization formalism (\ref{c})
 and microscopic description of the entropy of the black ring
 are in agreement with the entropy calculated by the macroscopic entropy function formalism
 (\ref{entropy}),
 \be
 S_{\rm mac}=S_{\rm mic}=2\pi Lp^2.\label{ss}
 \ee


 \subsection{Brown-Henneaux approach}\label{br}
 In this section we recalculate the microscopic entropy of
 supersymmetric black rings from another viewpoint  by using the Kerr/CFT formalism
 \cite{GHSS}.
 In this method, which is intrinsically a generalization of the Brown-Henneaux approach
 \cite{BH}, the Virasoro generators of the CFT dual are related to the  asymptotic
 symmetry group (ASG) of the near horizon metric.
 The asymptotic symmetry group (ASG) of a near horizon metric is the group of allowed
 symmetries modulo trivial symmetries. By definition, an allowed symmetry transformation
  obeys the specified  boundary conditions \cite{GHSS}.
 A possible boundary condition for the fluctuations around the geometry
 (\ref{met})   is,
 \be
 h_{\mu\nu}\sim\mathcal{O}\left(\begin{array}{ccccc}
 r^2&~ 1/r^2&~ 1/{r}&~ r &~ 1 \\
 ~&~{1}/{r^3}&~ {1}/{r^2}&~{1}/{r}^2&~{1}/{r}\\
 ~&~&~{1}/{r}&~{1}/{r}&~{1}/{r}\\
 ~&~&~&~1/r&~1\\
 ~&~&~&~&~1
 \end{array}\right),\label{h0}
 \ee
 in the basis $(t, r, \theta, \phi, \psi)$.
 It is easy to show that the general diffeomorphism preserving the boundary conditions (\ref{h0})
 is given by,
 \bea
 \zeta&=&\left[C+\mathcal{O}(\frac{1}{r^3})\right]\rd_t+\left[r\epsilon'(\psi)+\mathcal{O}(1)\right]\rd_r+
 \mathcal{O}(\frac{1}{r})\rd_\theta\nn\\
 &+&\mathcal{O}(\frac{1}{r^2})\rd_{\phi}+
 \left[\epsilon(\psi)+\mathcal{O}(\frac{1}{r^2})\right]\rd_{\psi},\label{br-solution}
 \eea
 where $C$  is an arbitrary constant and $\epsilon(\psi)$ is the arbitrary  smooth periodic functions of $\psi$.
 By using the basis $\epsilon_n(\psi)=-e^{-in\psi}$ for the function $\epsilon(\psi)$,
 it is easy to show that the ASG generates contains a Virasoro
 algebra generated by
 \bea
 \zeta_n=-e^{-in\psi}\rd_{\psi}-in~r~e^{-in\psi}\rd_r,\label{g-br}
 \eea
 which satisfy  $[\zeta_m , \zeta_n]=-i(m-n)\zeta_{n+n}$.

 The generator of a diffeomorphism has a conserved charge. The charges associated to the diffeomorphisms
 (\ref{g-br}) are defined by \cite{BB},
 \be
 Q_\zeta=\frac{1}{8\pi}\int_{\rd\Sigma}k_{\zeta}[h,g],\label{Q}
 \ee
 where $\rd\Sigma$ is spatial surface at infinity and
 \bea
 k_\zeta[h,g]&=&\frac{1}{2}[\zeta_\nu\nabla_\mu h-\zeta_\nu\nabla_\sigma h_\mu^{~\sigma}+\zeta_\sigma\nabla_\nu h_\mu^{~\sigma}+
 \frac{h}{2}\nabla_\nu\zeta_\mu\nn\\&-&h_\nu^{~\sigma}\nabla_\sigma\zeta_\mu
 +\frac{1}{2}h_{\nu\sigma}(\nabla_\mu\zeta^\sigma+\nabla^\sigma\zeta_\mu)]*(dx^\mu\wedge
 dx^\nu),
 \eea
  in which  $*$ denotes the Hodge dual in 5D.
 In the Brown-Henneaux  approach \cite{BH}
  the central charge is given by
 \be
 \frac{1}{8\pi}\int_{\rd\Sigma}k_{\zeta_m}[\mathcal{L}_{\zeta_n},g]
 =-\frac{i}{12}c(m^3- m)\delta_{{m+n},0}.\label{Qc}
 \ee
 Plugging the metric (\ref{met}) and diffeomorphisms (\ref{g-br})
 in (\ref{Qc}) one obtains,
 \be
 c=6p^3,\label{c-br}
 \ee
 which is in agreement with the c-extremization result (\ref{c}).

 The Frolov-Thorne temperature can be determined by identifying  quantum numbers
 of a matter field in the near horizon geometry with those in
 original geometry.
 For the chiral CFT given by (\ref{g-br}) a  matter field can be expanded in
 eigen modes of the asymptotic energy $\omega$
 and  angular momentum $m$ as
 \be
 \Phi=\sum_{\omega , m , l}\varphi_{\omega m l} e^{-i(\omega t-m\psi)}f_{l}(r,\theta, \phi),\label{Phi}
 \ee
 Similar to \cite{GHSS} one can show that, here,  the Frolov-Thorne
 temperature is
 \be
 T_{\rm FT}=\frac{1}{2\pi e_0}.\label{T}
 \ee
 The Cardy formula gives the microscopic entropy of  chiral
 CFT (\ref{g-br}) as follows,
 \be
 S_{\rm CFT}=\frac{\pi^2}{3}c~T_{\rm FT}=2\pi Lp^2,\label{ent-br}
 \ee
 which is in precise agreement with the result obtained by utilizing the entropy function and the c-extremization
 methods (\ref{ss}).

 \textbf{Acknowledgement}. We would like to thank M. Alishahiha, H. Ebrahim and R. Fareghbal for useful comments and discussions.

 \appendix

 \section{Killing vectors of AdS$_2\times S^1$ geometry}\label{Ap}

 In this appendix we derive the Killing vectors of AdS$_2\times S^1$ geometry
 which is appeared in the near horizon of black ring solution of $\mathcal{N}=2$ five dimensional supergravity.
 The metric of this part is (\ref{met0}),
 \be
 ds^2=-p^2(\frac{dr^2}{4r^2}+\frac{L^2}{p^2}d\psi^2+\frac{L r}{p} d\psi
 dt),
 \ee
 and Killing equation is
 \be
 X^{\rho}\rd_\rho g_{\mu\nu}+\rd_\mu X^\rho g_{\rho\nu}+\rd_\nu
 X^\rho g_{\mu\rho}=0.
 \ee
 The components of Killing equation are
 \bea
 &&\rd_tX^\psi=0,\label{k1}\\
 &&\rd_tX^r+\frac{2Lr^3}{p}\rd_rX^\psi=0,\label{k2}\\
 &&X^r+r\rd_t X^t+r\rd_\psi X^\psi=0,\label{k3}\\
 &&X^r-r\rd_rX^r=0,\label{k4}\\
 &&\rd_\psi
 X^r+\frac{2Lr^3}{p}\rd_rX^t+\frac{4L^2r^2}{p^2}\rd_rX^\psi=0,\label{k5}\\
 &&\rd_\psi X^t+\frac{2L}{p r}\rd_\psi X^\psi=0.\label{k6}
 \eea
 Equations (\ref{k1}) and (\ref{k4}) show that,
 \be
 X^\psi=f(r,\psi),\hspace{10mm}X^r=rg(t,\psi).
 \ee
 So we can simplify (\ref{k1})-(\ref{k6}) to obtain,
 \bea
 &&\rd_tg(t,\psi)+\frac{2Lr^2}{p}\rd_rf(r,\psi)=0,\label{kk2}\\
 &&g(t,\psi)+\rd_t X^t+\rd_\psi f(r,\psi)=0,\label{kk3}\\
 &&\rd_\psi
 g(t,\psi)+\frac{2Lr^2}{p}\rd_rX^t+\frac{4L^2r}{p^2}\rd_rf(r,\psi)=0,\label{kk5}\\
 &&\rd_\psi X^t+\frac{2L}{p r}\rd_\psi f(r,\psi)=0.\label{kk6}
 \eea
 In (\ref{kk2}) the first term is a function of $t$ and  $\psi$ and the second term is a function of
 $r$ and $\psi$. Therefore,  each term only is a function of $\psi$,
 \be
 \rd_tg(t,\psi)=-\frac{2Lr^2}{p}\rd_rf(r,\psi)=h(\psi),
 \ee
 and consequently,
 \be
 g_(t,\psi)=h(\psi)t+g_1(\psi),\hspace{10mm}f(r,\psi)=\frac{p}{2Lr}h(\psi)+f_1(\psi).\label{gf}
 \ee
 Now we can simplify (\ref{kk3})-(\ref{kk6}) as,
 \bea
 &&h(\psi)t+g_1(\psi)+\rd_t X^t+\frac{p}{2Lr}\rd_\psi h(\psi)+\rd_\psi f_1(\psi)=0,\label{kkk3}\\
 &&\rd_\psi
 h(\psi)t+\rd_\psi g_1(\psi)+\frac{2Lr^2}{p}\rd_rX^t-\frac{2L}{p r}h(\psi)=0,\label{kkk5}\\
 &&\rd_\psi X^t+\frac{1}{r^2}\rd_\psi h(\psi)+\frac{2L}{p r}\rd_\psi f_1(\psi)=0.\label{kkk6}
 \eea
 From (\ref{kkk3}) one finds
 \be
 X^t=-\left(\frac{1}{2}h(\psi)t^2+g_1(\psi)t+\frac{p}{2Lr}\rd_\psi h(\psi)t+\rd_\psi
 f_1(\psi)t\right)+I(r,\psi).\label{kkkk3}
 \ee
 (\ref{kkkk3}) and (\ref{kkk5}) give
 \be
 2\rd_\psi h(\psi)t+\rd_\psi g_1(\psi)+\frac{2Lr^2}{p}\rd_rI(r,\psi)-\frac{2L}{p r}h(\psi)=0.\label{kkkk5}\\
 \ee
 This implies that,
 \be
 \rd_\psi
 h(\psi)=0\hspace{10mm}\Rightarrow\hspace{10mm}h(\psi)=c_1.\label{h}
 \ee
 Thus (\ref{gf}) simplifies as
 \be
 g_(t,\psi)=c_1t+g_1(\psi),\hspace{20mm}f(r,\psi)=\frac{p}{2Lr}c_1+f_1(\psi),
 \ee
 and (\ref{kkk3})-(\ref{kkkk3}) become
 \bea
 &&c_1t+g_1(\psi)+\rd_t X^t+\rd_\psi f_1(\psi)=0,\label{kkkkk3}\\
 &&\rd_\psi g_1(\psi)+\frac{2Lr^2}{p}\rd_rX^t-\frac{2L}{p r}c_1=0,\label{kkkkk5}\\
 &&\rd_\psi X^t+\frac{2L}{p r}\rd_\psi f_1(\psi)=0,\label{kkkkk6}\\
 &&X^t=-\left(\frac{1}{2}c_1t^2+g_1(\psi)t+\rd_\psi
 f_1(\psi)t\right)+I(r,\psi).\label{xt}
 \eea
 By (\ref{xt}), Eqs.(\ref{kkkkk5}) and (\ref{kkkkk6})
 simplify to
 \bea
 &&\rd_\psi g_1(\psi)+\frac{2Lr^2}{p}\rd_rI(r,\psi)-\frac{2L}{p r}c_1=0,\label{kkkkkk5}\\
 &&-\left(\rd_\psi g_1(\psi)+\rd_\psi^2
 f_1(\psi)\right)t+\rd_\psi I(r,\psi)+\frac{2L}{p r}\rd_\psi f_1(\psi)=0.\label{kkkkkk6}
 \eea
 From (\ref{kkkkkk5}) one obtains
 \be
 I(r,\psi)=-\frac{1}{2r^2}c_1+\frac{p}{2Lr}\rd_\psi g_1(\psi)+I_1(\psi),
 \ee
 and so (\ref{kkkkkk6}) becomes
 \be
 -t\left(\rd_\psi g_1(\psi)+\rd_\psi^2
 f_1(\psi)\right)+\frac{p}{2Lr}\rd_\psi^2 g_1(\psi)+\rd_\psi I_1(\psi)+\frac{2L}{p r}\rd_\psi
 f_1(\psi)=0,
 \ee
 which implies that,
 \bea
 &&\rd_\psi g_1(\psi)+\rd_\psi^2 f_1(\psi)=0,\\
 &&\frac{p}{2L}\rd_\psi^2 g_1(\psi)+\frac{2L}{p}\rd_\psi
 f_1(\psi)=0,\\
 &&\rd_\psi I_1(\psi)=0.
 \eea
 Thus,
 \bea
 &&g_1(\psi)+\rd_\psi f_1(\psi)=c',\label{c2}\\
 &&\rd_\psi g_1(\psi)+\frac{4L^2}{p^2}
 f_1(\psi)=c_3,\label{c3}\\
 &&I_1(\psi)=c_4.
 \eea
 Now we can simplify our results.
 From (\ref{c2}) and (\ref{c3}) one obtains,
 \be
 -\rd_\psi^2f_1(\psi)+\frac{4L^2}{p^2}f_1(\psi)=c_3\Rightarrow
 f_1(\psi)=c_5e^{\frac{2L}{p}\psi}+c_6e^{-\frac{2L}{p}\psi}+\frac{p^2}{4L^2}c_3,
 \ee
 Therefore,
 \be
 g_1(\psi)=c_2+\frac{2L}{p}(c_6e^{-\frac{2L}{p}\psi}-c_5e^{\frac{2L}{p}\psi}).
 \ee
 Alternatively we can solve $g_1(\psi)$ as,
 \be
 -\rd_\psi^2g_1(\psi)+\frac{4L^2}{p^2}g_1(\psi)=\frac{4L^2}{p^2}c_2\Rightarrow
 g_1(\psi)=c_7e^{\frac{2L}{p}\psi}+c_8e^{-\frac{2L}{p}\psi}+c_2,
 \ee
 which is consistent with the first solution.
 $c_8$.

 In summary,
 \bea
 &&I(r,\psi)=-\frac{1}{2r^2}c_1-\frac{2L}{p r}\left(c_5e^{\frac{2L}{p}\psi}+c_6e^{-\frac{2L}{p}\psi}\right)+c_4,\\
 &&g(t,\psi)=c_1t+c_2+\frac{2L}{p}(c_6e^{-\frac{2L}{p}\psi}-c_5e^{\frac{2L}{p}\psi}),\\
 &&f(r,\psi)=\frac{p}{2Lr}c_1+c_5e^{\frac{2L}{p}\psi}+c_6e^{-\frac{2L}{p}\psi}+\frac{p^2}{4L^2}c_3.\\
 \eea
 So,
 \bea
 &&X^t=-\left(\frac{1}{2}c_1t^2+c_2t\right)
 -\frac{1}{2r^2}c_1-\frac{2L}{p r}\left(c_5e^{\frac{2L}{p}\psi}+c_6e^{-\frac{2L}{p}\psi}\right)+c_4,\\
 &&X^r=r\left(c_1t+c_2+\frac{2L}{p}(c_6e^{-\frac{2L}{p}\psi}-c_5e^{\frac{2L}{p}\psi})\right),\\
 &&X^\psi=\frac{p}{2Lr}c_1+c_5e^{\frac{2L}{p}\psi}+c_6e^{-\frac{2L}{p}\psi}+\frac{p^2}{4L^2}c_3.
 \eea
 Thus the killing vector expand as follows,
 \bea
 X=&X^t&\rd_t+X^r\rd_r+X^\psi\rd_\psi\nn\\
 =&-&c_1\left(\frac{1}{2}(t^2+r^{-2})\rd_t-rt\rd_r-\frac{p}{2Lr}\rd_\psi\right)\nn\\
 &-&c_2\left(t\rd_t-r\rd_r\right)
 +c_3\frac{p^2}{4L^2}\rd_\psi
 +c_4\rd_t\nn\\
 &-&c_5\frac{2L}{p}e^{\frac{2L}{p}\psi}\left(\frac{1}{r}\rd_t+r\rd_r-\frac{p}{2L}\rd_\psi\right)\nn\\
 &-&c_6\frac{2L}{p}e^{-\frac{2L}{p}\psi}\left(\frac{1}{r}\rd_t-r\rd_r-\frac{p}{2L}\rd_\psi\right),
 \eea
 and consequently, there are six  isometries generated by,
 \bea
 K_1&=&\frac{1}{2}(t^2+r^{-2})\rd_t-rt\rd_r-\frac{p}{2Lr}\rd_\psi,\hspace{5.5mm}K_2=t\rd_t-r\rd_r,\\
 K_3&=&\rd_\psi,\hspace{54.5mm}K_4=\rd_t\nn\\
 K_5&=&e^{\frac{2L}{p}\psi}\left(\frac{1}{r}\rd_t+r\rd_r-\frac{p}{2L}\rd_\psi\right),\hspace{11mm}
 K_6=e^{-\frac{2L}{p}\psi}\left(\frac{1}{r}\rd_t-r\rd_r-\frac{p}{2L}\rd_\psi\right).\nn
 \eea


\end{document}